\documentclass{article}
\usepackage{spconf,amsmath,graphicx}
\usepackage{booktabs}
\usepackage{multirow}
\usepackage{tikz}
\usepackage{pgfplots}
\usepackage{CJKutf8}
\usepackage{makecell}
\usepackage{hyperref}
\usepackage{cite}
\usepackage{caption}
\pgfplotsset{width=6cm, compat=1.4}
\title{Investigating Zero-Shot Generalizability on Mandarin-English Code-Switched ASR and Speech-to-text Translation of Recent Foundation Models with Self-Supervision and Weak Supervision}
\name{Chih-Kai Yang$^{*}$, Kuan-Po Huang$^{*\dagger}$\thanks{*Equal contribution.}, Ke-Han Lu, Chun-Yi Kuan, Chi-Yuan Hsiao, Hung-yi Lee}
\address{
$^{all}$National Taiwan University \quad $^{\dagger}$ASUS Intelligent Cloud Services \quad
}

\begin{document}
\ninept
\maketitle
\setlength{\abovedisplayskip}{3pt}
\setlength{\belowdisplayskip}{3pt}
\begin{abstract}
This work evaluated several cutting-edge large-scale foundation models based on self-supervision or weak supervision, including SeamlessM4T, SeamlessM4T v2, and Whisper-large-v3, on three code-switched corpora. We found that self-supervised models can achieve performances close to the supervised model, indicating the effectiveness of multilingual self-supervised pre-training. We also observed that these models still have room for improvement as they kept making similar mistakes and had unsatisfactory performances on modeling intra-sentential code-switching. In addition, the validity of several variants of Whisper was explored, and we concluded that they remained effective in a code-switching scenario, and similar techniques for self-supervised models are worth studying to boost the performance of code-switched tasks.

\end{abstract}
\vspace{-8pt}
\section{Introduction}
\vspace{-5pt}
\label{sec:intro}
Code-switching (CS) refers to the act of using multiple languages and switching the used languages during speaking or writing and is extremely common in countries having several official languages due to historical and cultural reasons. As it frequently occurs in daily conversations, speech technologies of it are strongly in demand for communication and connection around the world. Unfortunately, code-switching speech processing remains relatively unexplored, and it's still a pivotal challenge for the speech community. 

Over the past few years, there have been several works trying to develop technologies of code-switching for several tasks, such as automatic speech recognition (ASR)\cite{chi-bell-2022-improving, liu2023reducing}, speech-to-text translation (ST)\cite{weller-etal-2022-end}, etc. %However, most of the methods in these works were designed for a specific task and hence are not able to become task-universal. 
However, they are difficult to generalize to more languages or different kinds of tasks as they are mostly in a task-specific design and require high-quality labeled data that are hard to collect. Consequently, the further development is restricted.

% In the speech community, one popular way to resolve the issue of data scarcity is self-supervised learning, where a foundation model is first pre-trained on lots of unlabeled data, which are much easier to collect, to learn to encode the speech into faithful speech representations before fine-tuning on downstream tasks. The pre-training and the resulting foundation models\cite{baevski2020wav2vec, hsu2021hubert} have been shown effective in improving downstream performances with a relatively small amount of labeled data. Foundation models pre-trained on multilingual datasets are even shown to be 

In the speech community, there are two popular ways to resolve the issue of scarcity of high-quality labels. The first one is self-supervised learning, where a foundation model is first pre-trained on lots of unlabeled data, which is much easier to collect, to learn to encode the speech into faithful speech representations before fine-tuning on downstream tasks. The pre-training and the resulting foundation models\cite{baevski2020wav2vec, hsu2021hubert, wavlm} have been shown effective in improving downstream performances with a relatively small amount of labeled data. The other is weakly supervised learning, in which the requirement of label quality is relaxed and the model is trained on noisier datasets. This has been shown to be useful for ASR\cite{GigaSpeech2021, cheng21c_interspeech, tang-etal-2022-speechnet} when striking a balance between data quantity and quality.

Recently, via scaling up the amount of data and broadening the coverage of languages, large-scale multilingual models based on the two learning paradigms have been developed and publicly released\cite{seamlessv1, seamlessv2, mms, whisper}. These models demonstrate a strong capability of modeling multilingual speech and high proficiency in a variety of downstream tasks through multi-task training or delicate architecture design, making them universally applicable across the world. For some of them, their strength has been even further enhanced by techniques inspired by natural language processing, such as prompt-conditional fine-tuning\cite{clairaudience} and in-context learning\cite{SICLwhisper}. Therefore, considering the scarcity of code-switched data and the computation cost of fine-tuning these large-scale models, adopting these publicly available multilingual models in a zero-shot manner appears to be a more practical and efficient approach for code-switching scenarios. 

However, to our best knowledge, whether the validity of these models still holds when modeling code-switching remains unknown, since they were recently released and the related research is limited. To this end, we use several corpora that are either existent or collected by ourselves to evaluate the zero-shot generalizability of these models, as well as some of their variants, on Mandarin-English code-switched ASR and ST. 

The contributions of this paper are: (1) Exploring the limits of recent cutting-edge large-scale models based on self-supervision and weak supervision on code-switched ASR and ST,
%Providing investigation and comparison of recent cutting-edge large-scale models based on self-supervision and weak supervision in code-switching scenarios, 
(2) Pointing out the shortcomings of these models, including some error patterns due to their multi-task nature and the unsatisfactory performance on corpus entirely composed of intra-sentential CS, and (3) Verifying some variants of Whisper, e.g. prompt-conditional fine-tuning and speech-based in-context learning, can be generalized to boost the performance of Whisper on code-switched ASR and ST, even though they are not originally proposed for such a scenario, and encouraging researches that develop similar techniques for self-supervised models.

%\section{Related work}
%\label{sec:related}
%SeamlessM4T, a versatile multilingual system, is designed to accommodate both speech and text inputs or outputs. Its capabilities extend to diverse tasks, including speech-to-speech translation, speech-to-text translation, text-to-speech translation, text-to-text translation, and automatic speech recognition. 
%Additionally, the Massively Multilingual Speech (MMS) model, renowned for its extensive language support, demonstrates proficiency in multilingual speech recognition, speech synthesis, and speech language identification.

%\section{Investigated models and evaluation dataset}
\vspace{-8pt}
\section{Models and datasets}
\label{sec:zero-cs}
\vspace{-4pt}
\subsection{Investigated models}
\label{models}
%We selected several recent models with self-supervision or weak supervision that have been demonstrated to be powerful for multilingual ASR or speech-to-text translation tasks and investigated their task generalizability to tasks related to code-switching.
Our research involves a selection of models employing either self-supervision or weak supervision techniques. These models are notably effective in multilingual automatic speech recognition (ASR) and speech-to-text translation (ST) tasks. We focused on assessing the ability of these models in code-switching scenarios to better understand their zero-shot generalizability to code-switching tasks.
\vspace{-8pt}
\subsubsection{Self-supervised foundation models}
\label{self-supervised}
%\textbf{MMS and TCS}. MMS\cite{mms} is a recent multilingual speech foundation model whose pre-training data covers 0.5M hours of speech and 1406 languages. For multilingual ASR, MMS introduced language-specific adapters and LM heads for finetuning on each language, enabling parameter sharing of the foundation model across languages. During inference, MMS loads the corresponding adapter of the user-specified language for ASR. Though this approach was shown to be powerful and robust enough on multilingual ASR, using a single language-specific adapter turns out to be undoubtedly inadequate when it comes to code-switching ASR since the model needs to predict tokens from several languages instead of a single language. 
\textbf{MMS and TCS.} MMS~\cite{mms}, a recent foundational model for multilingual speech, has been pre-trained on an extensive dataset comprising 0.5M hours of speech across 1406 languages. For multilingual ASR, MMS employs language-specific adapters and language model (LM) heads, which allow for fine-tuning on individual languages while sharing parameters of the foundational model across different languages. During inference, MMS loads the corresponding adapter of the user-specified language for ASR. Though this approach was shown to be powerful and robust enough in multilingual ASR, using a single language-specific adapter turns out to be undoubtedly inadequate when it comes to CS ASR since the model needs to predict tokens from several languages instead of a single language.

To address this issue, Kulkarni et al.\cite{kulkarni2023adapting} proposed two approaches to employ the information from several language adapters. Among the two methods, Transformer Code Switching (TCS), in which the transformer is used to identify the CS boundaries and predict framewise binary code sequences to selectively utilize the output of the two adapters, was reported to perform better on Mandarin-English CS ASR. Hence, we chose TCS to be the reference level in our experiments for CS ASR. It should be noted that the TCS module has been fine-tuned with a small amount of code-switched data, and thus the CS ASR is not unseen for it.

\textbf{SeamlessM4T}. SeamlessM4T\cite{seamlessv1} is a multimodal, multi-task, and multilingual model that supports several tasks like ASR and ST. For the speech modality, it utilizes w2v-BERT 2.0, a self-supervised speech model that improves w2v-BERT \cite{w2vbert} with additional codebooks in its contrastive learning and masked prediction learning objectives as well as the introduction of the masked prediction task using random projection quantizers (RPQ) \cite{pmlr-v162-chiu22a}. The w2v-BERT 2.0 model was pre-trained on 1M hours of speech data, covering over 143 languages. The learned self-supervised representations then became one of the most important keys to the great performance of the model. In our experiments, we evaluate both the publicly available \textit{medium} and \textit{large} configurations of SeamlessM4T.

%\textbf{SeamlessM4T v2}. SeamlessM4T v2\cite{seamlessv2} is a further improved version of the aforementioned SeamlessM4T. Though many changes were made for the superior performance of SeamlessM4T v2, speaking of the self-supervised pre-training phase, the most influential improvement may be the scaling up of the pre-training dataset. SeamlessM4T v2 used 4.5M hours of speech data for pre-training, which is far more than the amount of data used in the previous version. This contributed to the significant performance improvement on various tasks.

\textbf{SeamlessM4T v2}. SeamlessM4T v2\cite{seamlessv2} is a further improved version of SeamlessM4T. Despite many changes being made for the superior performance of SeamlessM4T v2, the scaling up of the pre-training dataset contributes the most to the significant performance improvement during the self-supervised pre-training phase.
\vspace{-8pt}
\subsubsection{Weakly supervised foundation model and its variants}
\label{weak-supervised}
\textbf{Whisper}. Whisper\cite{whisper} is a series of transformer-based sequence-to-sequence models trained on a large-scale web-scraped speech dataset. One of its characteristics is the multi-task training for various tasks including multilingual ASR, X$\rightarrow$En ST, language identification, and timestamp prediction. Special tokens like the language tokens (e.g. $<$$|\mathtt{zh}|$$>$), task tokens (e.g. $<$$|\mathtt{transcribe}|$$>$), etc., are used as input of the Whisper decoder to specify the information of the task. As many Whisper models vary in size from 39M (tiny) to 1550M parameters (large), as well as in capabilities (multilingual or not), we choose to evaluate \textbf{Whisper-large-v3}, which may be the most powerful multilingual Whisper model that is trained on 5M hours of weakly labeled data and pseudo-labeled data.
%As there are many Whisper modes with different sizes, ranging from 39M (tiny) to 1550M parameters (large), and capabilities (multilingual or not), we choose to evaluate \textbf{Whisper-large-v3}, which may be the most powerful multilingual Whisper model that is trained on 5M hours of weakly labeled data and pseudo-labeled data.

\textbf{Variants of Whisper}. We also incorporate several works proposed to boost the performance of Whisper in different scenarios into our experiments to see whether they can improve Whisper on tasks related to code-switching. Peng et al.\cite{concat-prompt} proposed to provide Whisper with the concatenation of the language tokens of the languages occurring in the utterance (e.g. $<$$|\mathtt{zh}|$$>$$<$$|\mathtt{en}|$$>$) and showed that this can improve the performance of Whisper on CS ASR in a zero-shot manner. In the rest of the paper, we denote this method as $\mathtt{concat}$ and denote using a single language token of the matrix language of the corpora only (e.g. $<$$|\mathtt{zh}|$$>$) as $\mathtt{nonconcat}$. We examine whether this simple approach can be generalized to CS ST. 

Wang et al.\cite{SICLwhisper} showed that performing speech-based in-context learning (\textbf{SICL}) by providing the audio-transcription pairs as examples and a prompt with task information to Whisper model helps in the ASR task of Chinese dialects. We examine whether this methodology works in the CS scenario as well.

\textbf{Clairaudience}\cite{clairaudience} is a prompt-conditionally fine-tuned version of Whisper on the ASR task, in which a prompt consisting of the domain tags $\mathtt{domain\_N}$ with the format ``$\mathtt{domain:}$ $\mathtt{\{domain\_1\},}$ $\mathtt{\{domain\_2\}, ...}$" was used for finetuning. It was demonstrated to be effective in making Whisper more domain-sensitive and improving the performance on several domain-specific corpora. As Clairaudience was originally fine-tuned and evaluated on corpora that can be classified into concrete and specific domains (e.g. finance), we investigate whether the methodology remains effective when dealing with abstract and general properties of the dataset such as code-switching.

\vspace{-8pt}
\subsection{Evaluation datasets}
\label{dataset}
\vspace{-4pt}
We adopted three different corpora, including ASCEND \cite{lovenia2022ascend}, CSZS-correct \cite{huang2023zero}, and NTUML2021, to make the evaluation as comprehensive as possible. We briefly introduce these datasets.

\textbf{ASCEND} is a widely used Mandarin-English code-switched corpus with different topics of recorded conversational speech data, including education, technology, persona, philosophy, and sports. The training, validation, and testing splits contain 9.8K, 1.1K, and 1.3K utterances with a total duration of 8.78, 0.92, and 0.92 hours respectively. We adopted the testing split for evaluation. %We access the corpus via Hugging Face\footnote{\href{https://huggingface.co/datasets/CAiRE/ASCEND}{https://huggingface.co/datasets/CAiRE/ASCEND}} and adopt the testing split for evaluation.

\textbf{CSZS-correct} is an ASR corpus derived from a recently proposed code-switched benchmark\footnote{\href{https://github.com/nobel861017/cs_zs_baseline}{https://github.com/nobel861017/cs\_zs\_baseline}}~\cite{huang2023zero} originally designed to assess the linguistic aspects of speech encoders. The benchmark consists of paired synthesized utterances in which one utterance is judged erroneous by rigorous human annotations due to semantic inconsistency or grammatical unacceptability in its content and the other is judged correct. One of its features is that all the utterances in the benchmark contain intra-sentential CS and have extremely wide coverage of diverse topics and domains, making the benchmark more challenging than other commonly used corpora. We formed CSZS-correct by assembling the ``correct" subset from its Chinese-English track and the corresponding transcription to build an ASR corpus. The testing split of the resulting dataset contains 3176 utterances with a total duration of 4.1 hours.

\textbf{NTUML2021}\footnote{\href{https://huggingface.co/datasets/ky552/ML2021_ASR_ST}{https://huggingface.co/datasets/ky552/ML2021\_ASR\_ST}} is a speech corpus consisting of lecture recordings of the ``Machine Learning" course in 2021 at National Taiwan University along with the corresponding transcriptions and the translated English version of the transcriptions labeled by more than 20 bilingual native Chinese speakers, resulting in a Mandarin-English code-switching ASR and speech-to-text translation corpus. The base language of this corpus is Mandarin, with English serving as a guest language. Code-switching primarily occurs in discussions involving specific terminologies related to machine learning and deep learning. The domain-specific nature makes the NTUML2021 corpus difficult as the model may not have adequate domain knowledge to recognize and transcribe those terminologies accurately. Additionally, being the lecture recording, NTUML2021 is inherently noisy, which further adds to the difficulties. As the released corpus may be used for other purposes in the future, we divided it into training, validation, and testing splits, and only the testing split was used for ASR and ST in this paper. The testing split has 14916 utterances and a duration of 9 hours in total.

\vspace{-8pt}
\section{Experimental setup}
\label{sec:exp}
\vspace{-8pt}
\subsection{Metrics}
\label{metric}
Consistent with prior works, we chose the mixed error rate (MER), which treats Chinese characters as individual words, and the BLEU score as the evaluation metrics for CS ASR and ST, respectively.

In addition, we first processed the ground truths and the model predictions before the evaluation in the following way: (1) Since the transcriptions of NTUML2021 are in traditional Chinese and English, we first ensured all the Chinese characters in the ground truths and model predictions were converted into simplified Chinese, and lowercased all the English words as well, (2) punctuation was removed, and (3) spaces were inserted between Chinese characters and English words. 
%For ASR, we treat the Chinese characters as individual words and report the mixed error rate (MER) as the evaluation metric. Since the transcriptions of NTUML2021 are in traditional Chinese and English, we first ensured all the Chinese transcriptions were converted into simplified Chinese, and lowercased all the English words as well. In addition, punctuation was removed and spaces were inserted between Chinese characters and English words. 
%The MER performances were then computed through the JiWER library. Similar processing was conducted for ST as well.%the translation was modified into all lower cases and stripped of punctuation. 

%Consistent with standard practices, the BLEU score was employed as the primary metric for ST evaluation.
%As for the speech-to-text translation task, the above post-processing was conducted, and the BLEU score was reported as usual. 
% \footnote{\href{https://pypi.org/project/OpenCC/}{https://pypi.org/project/OpenCC/}}
% \footnote{\href{https://pypi.org/project/jiwer/}{https://pypi.org/project/jiwer/}}
\vspace{-10pt}
\subsection{Domain tags for Clairaudience}
\label{tag}
As mentioned in Sec. \ref{weak-supervised}, Clairaudience uses domain tags to make the model more sensitive and consistent with the domain information when transcribing. Therefore, we adopted a strategy similar to \cite{clairaudience} for generating domain tags. We used ``code-switching" as the common domain tag for all three datasets and dataset-specific domain tags were provided to the model as well for some corpora. Please note that as the transcriptions of the testing data won't be accessible in practice, we didn't deliberately generate the domain tags at the utterance level if they are not inherently provided in the datasets since this requires data transcriptions, though it's technically feasible.

In ASCEND, each audio clip is accompanied by its topic label (listed in Sec.\ref{dataset}). Therefore, besides ``code-switching", we used these associated topic labels as the additional dataset-specific domain tags. For NTUML2021, though the information of each utterance is absent, we included ``machine learning" in the prompt based on the prior knowledge of the corpus. Finally, as the covered topics of CSZS-correct are highly diverse and the metadata is unavailable, we didn't use dataset-specific domain tags for this corpus, and ``code-switching" was the only provided tag.
\vspace{-8pt}
\subsection{SICL and example selection}
\label{example}
Our implementation of SICL essentially followed the approach in \cite{SICLwhisper}. The audio input of Whisper was the concatenation of the example audio and the testing audio. The full stop symbol ``\begin{CJK}{UTF8}{gbsn}。\end{CJK}" in written Chinese was intentionally added at the end of the example transcription if the transcription did not end with it originally. The processed transcription was given to the Whisper decoder as a prefix. %\footnote{\href{https://github.com/openai/whisper/discussions/117}{https://github.com/openai/whisper/discussions/117}}. 
Our preliminary experiments indicated that including the full stop could significantly mitigate the hallucination, where Whisper tended to generate possible following words of the prefix, even though these words didn't exist in the input speech. 

One of the differences between the implementation of \cite{SICLwhisper} and ours is the number of examples. Since most data samples in our evaluation corpora are the audio clips of entire sentences instead of words in \cite{SICLwhisper}, we only used one example at a time to prevent the input from becoming too long, which might hurt the performance and make the assessments unfair. 

The other difference between our implementation and \cite{SICLwhisper} is the example selection. The strategy in \cite{SICLwhisper} is based on kNN, which needs an additional dataset to serve as the example pool for retrieval. However, in real-world applications, such a large example pool is typically unavailable. Hence, to simulate this situation, we only randomly sampled 10 CS audio-transcription pairs from the training split of each corpus as example candidates. An additional constraint of topic balancing was posed during the sampling of ASCEND to ensure the examples were not biased toward some specific topics. 

As for the prompt, we simply used ``This is a code-switching sentence. Transcribe it." or ``This is a code-switching sentence. Translate it." for Whisper-large-v3 depending on the encountered task and domain tags described in \ref{tag} with the format mentioned in \ref{weak-supervised} for Clairaudience. These prompts are denoted as $\mathtt{prompt}$ in the rest of the paper and may be used alone without SICL.

Each individual sampled pair was then provided to the model, accompanied by the associated $\mathtt{prompt}$, as a SICL example. The resulting performance was recorded, and the average of the top five performances out of the 10 examples was reported.

\vspace{-8pt}
\section{Results}
\label{sec:results}
\begin{table*}[ht]
\setlength\tabcolsep{5.0 pt}
\renewcommand{\arraystretch}{0.2}
\caption{MER(\%) of the investigated models on ASCEND, CSZS-correct, and NTUML2021. In the table, $\mathtt{concat}$ and $\mathtt{nonconcat}$ refer to using the concatenation of language tokens and using single language tokens, respectively. SICL and $\mathtt{prompt}$ represent the speech-based in-context learning and the prompts described in \ref{example} for simplicity. The best performance values in each block are marked in bold. %In each block, the numbers in bold represent the best performance on the corresponding corpora among the models.
}
\centering
\begin{tabular}{clccccc}
\toprule
        &   & & CSZS-correct           & ASCEND  & NTUML2021 & Avg\\
Index & Model &  Prompting strategy  &    MER($\downarrow$)  & MER($\downarrow$) & MER($\downarrow$) & MER($\downarrow$) \\
\midrule
\midrule
\multicolumn{7}{c}{Self-supervised foundation models}\\
\midrule
1 & TCS & -             & 70.59       & 47.71 & 73.83 & 64.04 \\
2 & SeamlessM4T \textit{medium} & - & 71.12 & 40.94 & 23.11 & 45.06 \\
3 & SeamlessM4T \textit{large} & - & 69.52 & 38.39 & 23.43 & 43.78 \\
4 & SeamlessM4T v2 & - & \textbf{63.69} & \textbf{25.86} & \textbf{14.26} & \textbf{34.60} \\
\midrule
\midrule

\multicolumn{7}{c}{Weakly supervised model}\\
\midrule
5 & Whisper-large-v3 & $\mathtt{nonconcat}$ & 60.17 & 30.29 & 10.11 & 33.52\\

\midrule
\midrule
\multicolumn{7}{c}{Variants of Whisper}\\
\midrule
6 & Whisper-large-v3 & $\mathtt{prompt}$ + $\mathtt{nonconcat}$ & 29.85 & 25.20 & \ \ 9.95 & 21.67\\
7 & Whisper-large-v3 & $\mathtt{concat}$ & 26.76 & 21.93 & 10.10 & 19.60\\
8 & Whisper-large-v3 & $\mathtt{prompt}$ + $\mathtt{nonconcat}$ + SICL & 13.56 & 13.53 & \ \ 9.72 & 12.27\\
9 & Whisper-large-v3 & $\mathtt{prompt}$ + $\mathtt{concat}$ + SICL & 13.80 & 13.95 & 11.58 & 13.11\\
10 & Clairaudience & $\mathtt{prompt}$ + $\mathtt{nonconcat}$ & 16.63 & 14.93 & 10.42 & 13.99\\
11 & Clairaudience & $\mathtt{prompt}$ + $\mathtt{nonconcat}$ + SICL & 13.41 & 13.49 & \textbf{\ \ 9.50} & 12.13 \\
12 & Clairaudience & $\mathtt{prompt}$ + $\mathtt{concat}$ & 15.81 & 26.86 & 22.03 & 21.57 \\
13 & Clairaudience & $\mathtt{prompt}$ + $\mathtt{concat}$ + SICL & \textbf{13.15} & \textbf{13.20} & \ \ 9.96 & \textbf{12.10}\\

\bottomrule
\end{tabular}
\label{tab:mer}
\vspace{-12pt}
\end{table*}

\begin{table}[ht]\scriptsize
\setlength\tabcolsep{2 pt}
\renewcommand{\arraystretch}{0.2}

\caption{BLEU score of the investigated models on the NTUML2021 speech-to-text translation task, with GPT-3.5 performing text-to-text translation on the transcriptions included as a topline.} %The $\mathtt{prompt}$ notation refers to ``This is a code-switching sentence. Translate it."}
\centering

\begin{tabular}{clcc}
\toprule
% & & NTUML2021 \\
Index & Model & Prompting strategy & BLEU ($\uparrow$) \\
\midrule
\midrule
14 & GPT-3.5 (text translation top-line) & - & 41.39 \\
\midrule
\midrule
\multicolumn{4}{c}{Speech-to-text translation (ST)}\\
\midrule
15 & SeamlessM4T v2 (self-supervised) & - & 22.46 \\
\midrule
16 & Whisper-large-v3 & $\mathtt{nonconcat}$ & 29.70 \\
17 & Whisper-large-v3 & $\mathtt{concat}$ & 29.79 \\
18 & Whisper-large-v3 & $\mathtt{prompt}$ + $\mathtt{nonconcat}$ & 29.97 \\
19 & Whisper-large-v3 & $\mathtt{prompt}$ + $\mathtt{concat}$ & 30.04 \\
20 & Whisper-large-v3 & $\mathtt{prompt}$ + $\mathtt{nonconcat}$ + SICL & \textbf{31.57} \\
21 & Whisper-large-v3 & $\mathtt{prompt}$ + $\mathtt{concat}$ + SICL & 31.49 \\
22 & Clairaudience & $\mathtt{prompt}$ + $\mathtt{nonconcat}$ & 7.23\\
23 & Clairaudience & $\mathtt{prompt}$ + $\mathtt{concat}$ & 7.23\\
24 & Clairaudience & $\mathtt{prompt}$ + $\mathtt{nonconcat}$ + SICL & 8.24\\
25 & Clairaudience & $\mathtt{prompt}$ + $\mathtt{concat}$ + SICL & 8.24\\
\bottomrule
\end{tabular}
\label{tab:st}
\vspace{-20pt}
\end{table}

\vspace{-5pt}
Our results on CS ASR are listed in Table \ref{tab:mer}. Extended evaluation results of SeamlessM4T v2, Whisper-large-v3, and its variants on NTUML2021 ST are in Table \ref{tab:st}.
\vspace{-7pt}
\subsection{Self-supervision and weak supervision}
We first discuss the performances of self-supervised models (exp. 1-4 and 15) and the original Whisper-large-v3 (exp. 5 and 16) without any variants.
As shown in Table \ref{tab:mer}, the most performant self-supervised model for CS ASR is SeamlessM4T v2, which benefits from both large-scale pre-training data (4.5x of those of SeamlessM4T and 9x of those of MMS) and the relatively abundant Mandarin ASR data (1.3x of those of SeamlessM4T and maybe 13x of those of MMS according to the estimation in \cite{kulkarni2023adapting}). This suggests that data scaling is far more crucial than the wide coverage of languages, although the latter has been shown to help capture the semantic and syntactic linguistic features of CS speech\cite{huang2023zero}.

It should be pointed out that despite being inferior to Whisper-large-v3 on ST (comparing exp. 15 and 16), SeamlessM4T v2 turns out to achieve performances that are quite close to those of the original Whisper-large-v3 (exp. 5 in Table \ref{tab:mer}) on CS ASR, even though the latter probably used more than 2x of Mandarin ASR data and extremely more English ASR data than the former\footnote{The data statistics in \cite{whisper} likely represent the minimum training data for Whisper-large-v3, though not officially confirmed.}. This hints that it's still possible for models with self-supervision that learn to extract high-quality speech representations to achieve comparable performance to those with (weak) supervision while using relatively limited labeled data, making self-supervised pre-training a practical approach when the labeled resources are scarce and expensive.
%Detailed data statistics are not formally reported, but the amount reported in \cite{whisper} can be reasonably assumed to be the lower bound of the amount of the training data used by Whisper-large-v3.

However, from Table \ref{tab:mer}, both self-supervised models and weakly supervised models obtain unsatisfactory performances on CSZS-correct. This may indicate that intra-sentential CS is difficult for these models probably due to the requirement of cross-lingual understanding, and there is still room for improvement.

We observed common error patterns on CS ASR made by SeamlessM4T v2 and Whisper-large-v3. First, the two models tended to spontaneously translate part of the speech into the other language when transcribing, which is unacceptable for ASR. This likely came from their multi-task nature. Furthermore, we also found that the two models tended to make mistranscriptions in some domain-specific terminologies.  There is no surprise since unlike the pre-training of LLM, the training in speech, no matter self-supervised pre-training or (weak) supervised training on downstream tasks, is not adequate to equip the models with world knowledge due to the much more complex nature of speech modality compared with text. Examples of the error patterns are demonstrated in Table \ref{tab:pattern}.

We also noticed that TCS performed much worse than other models on NTUML2021 (exp. 1), though it has been fine-tuned on code-switched data. The possible reason is two-fold. First, TCS is based on MMS ASR, which is estimated to be trained on only 1K hours of Mandarin speech\cite{kulkarni2023adapting}. Thus, the model may not be strong enough in Mandarin and likely has weak generalizability to Mandarin-English code-switching. In addition, as NTUML2021 is mostly in Chinese, this corpus is probably more difficult than the other two corpora for TCS, resulting in a relatively poor performance of TCS on NTUML2021 than the performances of other models.

\vspace{-8pt}
\subsection{Effectiveness of variants of Whisper}
We discuss the effectiveness of variants of Whisper on CS ASR and ST. First of all, comparing exp. 5 and 6, 16 and 18, 17 and 19, we observed that simply prompting can already bring improvement in average MER and BLEU. For CS ASR, the three variants (exp. 7 for Peng et al.\cite{concat-prompt} utilizing Mandarin and English language tokens, exp. 8 for Wang et al.\cite{SICLwhisper} combining prompts and SICL, exp. 10 for Liao et al.\cite{clairaudience} adopting prompt-conditionally fine-tuned Whisper with domain tags as prompts) are also effective. Specifically, among exp. 7, 8, 10, exp. 8 
achieved the best performance of CS ASR among the three approaches probably because it's the most direct way to demonstrate what code-switching is to Whisper. For CS ST, a similar trend is observed in Table \ref{tab:st}. Exp. 20 shows that performing SICL with prompts achieves the highest BLEU score on the NTUML2021 ST task. As for Clairaudience (exp. 22-25), since it is only fine-tuned on ASR, the performances on ST were relatively poor compared to other models. In addition, it's not sensitive to $\mathtt{concat}$, and SICL can only improve the performance a little bit.  
% and hence, is expected to perform poorly on the CS ST task.

%Exp. 6, 7, 8, and 10 in Table \ref{tab:mer} represent the performance of simply prompting Whisper-large-v3 and methods proposed in Peng et al.\cite{concat-prompt}, Wang et al.\cite{SICLwhisper} and Liao et al.\cite{clairaudience} on CS ASR respectively. As shown, simply prompting (exp. 6) can already bring a huge improvement in average MER, and all three methods (exp. 7, 8, 10) are more effective in zero-shot generalization to CS ASR, as they outperformed vanilla Whisper-large-v3 (exp. 5) and prompted Whisper-large-v3 (exp. 6) in most cases. Specifically,  Wang et al.\cite{SICLwhisper} (SICL accompanied by $\mathtt{prompt}$) boosted the performance the most among the three approaches probably because it's the most direct way to demonstrate what code-switching is to Whisper.

\begin{table}[ht]\scriptsize
\setlength\tabcolsep{4 pt}
\renewcommand{\arraystretch}{0.1}

\caption{Examples of error patterns of SeamlessM4T v2 and Whisper-large-v3. ``Auto translation" refers to spontaneously translating parts of speech into the other language. ``Wrong terminology" refers to mistranscribing domain-specific terminologies.}
\centering

\begin{tabular}{cccc}
\toprule
Pattern & Ground truth & SeamlessM4T v2 & Whisper-large-v3 \\
\midrule
\multirow{6.5}{*}{\makecell[c]{Auto \\translation}} & \begin{CJK}{UTF8}{gbsn} \makecell[c]{会关系到一个\\passenger的\\一个性性命} \end{CJK} & \begin{CJK}{UTF8}{gbsn} \makecell[c]{会关系到一个\\乘客的一个性命} \end{CJK} & \begin{CJK}{UTF8}{gbsn} \makecell[c]{会关系到一个\\客人的性命} \end{CJK} \\

\cmidrule{2-4}

 & \begin{CJK}{UTF8}{gbsn} \makecell[c]{来hong kong的时候\\是做那个} \end{CJK} & \begin{CJK}{UTF8}{gbsn} \makecell[c]{来香港的时候\\是做那个} \end{CJK} & \begin{CJK}{UTF8}{gbsn} \makecell[c]{来香港的时候\\是做那个} \end{CJK} \\ 

\midrule

\multirow{45}{*}{\makecell[c]{Wrong \\ terminology}} & \begin{CJK}{UTF8}{gbsn} \makecell[c]{然后我再把\\autoencoder\\的部分补完} \end{CJK} & \begin{CJK}{UTF8}{gbsn} \makecell[c]{然后我再把\\posenco\\的部分补完} \end{CJK} 
 & \begin{CJK}{UTF8}{gbsn} \makecell[c]{然后再把\\hotel call\\的部分补完} \end{CJK} \\

\cmidrule{2-4}

  & \begin{CJK}{UTF8}{gbsn} \makecell[c]{伞蛱蝶属 is a \\genus in the \\subfamily of \\lycaenidae} \end{CJK} 
  & \begin{CJK}{UTF8}{gbsn} \makecell[c]{散甲蝶属 its \\a genus in \\ the subfamily \\alyssinidae} \end{CJK}
  & \begin{CJK}{UTF8}{gbsn} \makecell[c]{三家蝶属 \\ lysagenous \\and subfamily of \\lycaemidae} \end{CJK} \\

\bottomrule
\end{tabular}

\label{tab:pattern}
\vspace{-20pt}
\end{table}

We further discuss the results of the combination of these approaches (exp. 9 and 11-13 in Table \ref{tab:mer} and exp. 21 in Table \ref{tab:st}). We found that $\mathtt{concat}$ had bad chemistry with the other two approaches, as the combination of $\mathtt{concat}$ and any one of the other approaches usually led to performance worse than using that approach alone (comparing exp. 8 with 9, 10 with 12, and 20 with 21), and it seems to only provide a marginal advantage when the three approaches were applied at the time (comparing exp. 11 with 13) for CS ASR, although it's the only one specifically proposed for CS scenario. Nevertheless, combining all three methods was still empirically the most effective way among the investigated methods for the zero-shot Mandarin-English CS ASR task.

From this observation, though it's just a budding understudied research direction\cite{hsu2023exploration, huang2023dynamicsuperb} for now, we believe the self-supervised speech foundation models can become more powerful shortly if similar techniques for them can be developed in depth.

%\vspace{-8pt}
%\subsection{Speech-to-text translation}
%Table \ref{tab:st} shows the results of evaluating Whisper on the NTUML2021 ST task. By comparing exp. 15 and 16, it can be seen that the original Whisper model outperforms self-supervised learning methods. By comparing exp. 16 and 18, 17 and 19, it is demonstrated that giving prompts during inference still improves the BLEU to a small extent.
%By comparing exp. 17 and 21, it is shown that SICL is also effective in improving translation performance when both the Mandarin and English language token is given. 
%given the fact that the performance of Whisper is already close to the text translation top-line generated by GPT-3.5.
\vspace{-10pt}
\section{Conclusion}
\vspace{-8pt}
\label{sec:conclusion}
%Evaluating the effectiveness of large-scale multilingual models based on self-supervised and weakly supervised models in CS ASR and ST tasks reveals that techniques like prompting and speech-based in-context learning can enhance performance in CS contexts. This research underscores the need for further development of models specifically tailored for CS, highlighting the importance of more nuanced approaches in multilingual speech technology.
This paper investigates the abilities of several cutting-edge self-supervised models, a weakly supervised model, and its variants on Mandarin-English code-switched ASR and ST. Our results show that self-supervised models can demonstrate comparable performance to the weakly supervised model, indicating the efficacy of the pre-training. We also point out some shortcomings of these models in modeling code-switching, especially intra-sentential code-switching. Techniques like in-context learning are shown to be helpful for the weakly supervised model in generalization, highlighting the necessity of similar ones for self-supervised models.
\bibliographystyle{IEEEbib}
\bibliography{strings,refs}

\end{document}